\documentclass[preprint]{aastex}

\usepackage{mathtext,bbm,amsmath,amsfonts,amssymb,indentfirst,syntonly,graphicx}
\usepackage[english]{babel}
\usepackage{slashbox}
\usepackage{calc}
\usepackage{tikz}
\usepackage{latexsym,graphicx,bbm}

\newcommand{\beq}{\begin{eqnarray}}
\newcommand{\eeq}{\end{eqnarray}}

\shorttitle{Band spectra of GRB in the prompt phase}
\shortauthors{Jiang et al.}

\begin{document}

\title{Gamma-ray polarization induced by cold electrons via Compton processes}

\author{ Zhe Chang\altaffilmark{1,2,*}, Yunguo Jiang\altaffilmark{1,2,\dag}, Hai-Nan Lin\altaffilmark{1,\ddag}}
\affil{\altaffilmark{1}Institute of High Energy Physics\\Chinese Academy of Sciences, 100049 Beijing, China}
\affil{\altaffilmark{2}Theoretical Physics Center for Science Facilities\\Chinese Academy of Sciences, 100049 Beijing, China}
\altaffiltext{*}{changz@ihep.ac.cn}
\altaffiltext{\dag}{jiangyg@ihep.ac.cn}
\altaffiltext{\ddag}{linhn@ihep.ac.cn}

\begin{abstract}
The polarization measurement is an important tool to probe the prompt emission mechanism in gamma-ray bursts (GRBs). The synchrotron photons
can be scattered by cold electrons in the outflow via Compton scattering processes. The observed polarization  depends
on both  the photon energy and the viewing angle. With the typical bulk Lorentz factor $\Gamma \sim 200$, photons with energy $E>10$ MeV
tend to have smaller polarization
 than photons with energy $E<1$ MeV. At the right viewing angle, i.e. $\theta \sim \Gamma^{-1}$, the polarization achieves its maximal value,
and the polarization angle  changes $90^{\circ}$ relative to the initial polarization direction. Thus, the synchrotron radiation
plus Compton scattering model can naturally explain  the $90^{\circ}$  change of the polarization angle in GRB 100826A.
\end{abstract}


\section{Introduction}
 Although a remarkable advance of investigations on GRBs was made in the past decades, the mechanism of the prompt emission of GRBs
is still unclear. The optically thin synchrotron radiation (SR) is believed to be the most promising mechanism to produce a broken
power law  spectrum, i.e. the empirical Band function \citep{Band:1993}.  One supporting evidence is from the ``Amati relation",
which states that the peak energy of the $\nu F_\nu$ spectrum $E_{\rm peak}$ is correlated with the isotropic equivalent radiation energy
$E_{\rm iso}$ \citep{Amarti:2002,Amarti:2006}. However, there are still some disputes in the low energy regime, namely the prediction
 of the SR  in the fast cooling phase $\alpha \sim -3/2$ is against a large number of the observed samples $\alpha \sim -1$ \citep{Preece:2000}.
 Essentially, the inverse Compton (IC) process combined with the SR will change
 the distribution of the  electrons and influence the spectra. \citet{Daigne:2011} also indicated that  $\alpha$ in the marginally
 fast cooling case can be up to $-2/3$, and the majority of the observed GRB prompt phase can be reconciled with the synchrotron origin.
 The SR process produces polarized light beams, the polarimetric observation of the prompt emission will provide us more information
beyond the spectrum and the light curve. The time lags between GeV and MeV photons in GRBs were used to constraint
 the Lorentz invariance violation (LIV) effects \citep{Chang:2012a}, and the polarization observation can
be used to constraint the CPT violation effects \citep{Toma:2012}. Many issues of polarizations in GRBs were investigated both theoretically and experimentally.

Theoretically, the gamma-ray polarization depends on both the emission mechanism and the geometry of the outflow. For synchrotron origin in the prompt phase,
 the maximal polarization is $\Pi_{\rm max}=(p+1)/(p+7/3)$, if the spectrum of electrons has a power-law index $p$ and the magnetic field
 is uniform at large scale in the plasma \citep{Rybicki:1979}. The configuration of the magnetic field is essential to produce the polarization.
 \citet{Granot:2003} indicated that an ordered magnetic field can produce $\Pi \geq 50 \%$ easily, and the  magnetic field component orthogonal to
the moving direction of the outflow produce  smaller $\Pi$ than the one parallel to the moving direction. \citet{Lyutikov:2003} calculated the stokes
 parameters and derived the pulse-averaged polarization with a given toroidal magnetic field. They showed that $\Pi \sim 56\%$ can be achieved for viewing
angles larger than $1/\Gamma$ and the observed maximum polarization is smaller than that due to the relativistic kinematic effect.
\citet{Nakar:2003} showed that the SR from random magnetic fields can lead to $\Pi \sim 30-35\%$, while a uniform field produces $\Pi \sim 45-50 \%$.
 However, a significant polarization can be obtained if the line of sight is within the small jet open angle $\theta_{\rm jet} \sim \Gamma^{-1}$.

The synchrotron origin is widely accepted as the successful theory to explain the GRB afterglows \citep{Sari:1998}.
The polarization in the afterglow phase has also been studied by several works \citep{Gruzinov:1999a,Gruzinov:1999b,Sari:1999}.
\citet{Gruzinov:1999a} showed  $\Pi\cong \Pi_{\rm max}/ \sqrt{N}$, where $N$ is the number of the magnetic field patches  in the visible region.
\citet{Sari:1999} investigated the polarization on the averaged emission site in the afterglow, and showed that the observed $\Pi$ is not likely to
exceed $20\%$. Considering the dynamics of the jet, the polarization direction will change $90^\circ$ before and after the jet breaking
time for an observer  moving away from the jet center. Although observations of the optical afterglow of  GRB 021004 and  GRB 020405 showed a
possible polarization evolution in both the direction and the degree, the cosmic dust contribution ($\sim 1\%$) can not be unambiguously
excluded \citep{Lazzati:2003,Covino:2003}. A recently observed GRB 091018  confirmed the polarization evolution theory \citep{Wiersema:2012}.
 \citet{Rossi:2004} studied the jet structure effects on the polarization in the afterglow.

As stated above, the IC processes are essential to understand the prompt emission mechanism. The IC induced polarizations were also
studied by several authors \citep{Shaviv:1995,Lazzati:2004,Eichler:2003,Levinson:2004,Toma:2009}. \citet{Shaviv:1995} showed that the IC scattering can
explain many features of the GRBs, such as the temporal features, multi-peak light curves, power-law spectra in the high energy regime, and also
the polarization prediction. \citet{Lazzati:2004} discussed the IC induced polarization in the point source limit, and indicated that $\Pi$ can be large
after the proper angular integration when a certain special geometry of the jet is realized. \citet{Toma:2009} used the Monte Carlo methods
to distinguish GRB polarizations produced in three different scenarios: the SR with a global ordered magnetic field (SO) model, the SR with
 random magnetic fields  model and the Compton drag  model. The simulation showed that the Compton drag model is favored when $\Pi>0.8$, and the SR with ordered
magnetic fields model is favored when $\Pi \sim 0.2-0.7$.

The early age observations of the polarization concerned mainly the optical and radio bands, and the typical value of $\Pi$ is
less than $10\%$ for many GRBs in the afterglow phase \citep{Taylor:1998,Frail:1998,Hjorth:1999,Wijers:1999,Bersier:2003,Greiner:2003,Steele:2009}.
 The first large linear polarization in the prompt emission was reported by {\it RHESSI} from GRB 021206 \citep{Coburn:2003}, although this result was challenged
 by  other independent groups \citep{Rutledge:2004, Wigger:2004}. The recent detection of the polarization in the prompt phase indicated
very large polarizations in GRB 041219A, GRB 100826A, GRB 110310 and GRB 110721A \citep{McGlynn:2007,Kalemci:2007,Gotz:2009,Yonetoku:2011,Yonetoku:2012,Toma:2012}. We
give a summary of these results in Table \ref{tab}. From the data, one can see that it is possible that the polarization in the prompt phase is larger than $80\%$,
which is beyond the maximal value of the SR induced polarization. Thus, the SR plus Compton scattering (CS) model is worth investigating. We wish that
 the initial polarization of SR photons and CS effect can not only supply a reasonable framework for observed GRB polarization, but also presents a consistent
scenario for the spectra, the light curves, and the polarizations.

The present paper is organized as follows. In Section \ref{sec:setup}, we  introduce the SR plus CS model, derive the polarization formula in
the observer frame, and show the characteristics of the model. In Section \ref{sec:casestudy}, we use this SR and CS effect combined model to discuss four GRB cases.
The conclusion and remarks are given in Section \ref{sec:con}.

\begin{deluxetable}{cccccc}
\tablecaption{Polarizations in observation \label{tab}}
%\tabletypesize{\scriptsize}
\tablewidth{0pt}
\tablehead{\colhead{GRB}&\colhead{Redshift}&\colhead{Energy Band}&\colhead{Polarization}&\colhead{Phase}&\colhead{References}}
\startdata
GRB 980329 & $\sim 3.5$ & radio & $<21$ \% $(2\sigma)$& afterglow & [1] \\
%GRB 980703 & 0.966 & radio & $<8.0$ \%& & \citet{Frail:1998} \\
GRB 990123 & 1.61 & optical & $<2.3$ \%& afterglow & [2]  \\
GRB 990510 & 1.619 & optical & $<1.7$ \% & afterglow& [3], [4] \\
GRB 020405 & 0.695 & optical & $<10$ \% & afterglow& [5] \\
GRB 021206  & / & $\gamma$-ray & $< 100$ \% & prompt & [6],[7]  \\
GRB 030329 & 0.168 & optical & $0.3 \sim 2.5$ \% & afterglow& [8], [9] \\
GRB 041219A & $\sim 0.3$ & $\gamma$-ray & $98\pm 33$ \%& prompt & [10], [11]  \\
GRB 090102 & 1.547 & optical & $10\pm 1$ \%& afterglow & [12]  \\
GRB 091208B & 1.063 & optical & $10.4\pm2.5 $ \%& afterglow &[13]   \\
GRB 100826A & / & $\gamma$-ray & $27\pm 11$  \% (2.9$\sigma$)&prompt &  [14]  \\
GRB 110301A & / & $\gamma$-ray & $70\pm 22$  \% (3.7$\sigma$)& prompt&[15]  \\
GRB 110721A & 0.382\tablenotemark{a} & $\gamma$-ray & $88_{-28}^{+16}$ \% (3.3$\sigma$)&prompt & [15], [16]\\
\enddata
\tablenotetext{a}{There is a caution that the afterglow observations of GRB 110721A were inconclusive, another candidate redshift is 3.512 \citep{Berger:2011}.}
\tablerefs{[1] \citet{Taylor:1998}; [2] \citet{Hjorth:1999}; [3] \citet{Covino:1999}; [4] \citet{Wijers:1999}; [5] \citet{Bersier:2003};
[6] \citet{Rutledge:2004}; [7] \citet{Wigger:2004}; [8] \citet{Greiner:2003}; [9] \citet{Caldwell:2003}; [10]  \citet{McGlynn:2007};
[11] \citet{Kalemci:2007};  [12] \citet{Steele:2009};
[13] \citet{Uehara:2012}; [14] \citet{Yonetoku:2011}; [15] \citet{Yonetoku:2012} [16] \citet{Berger:2011}.}
%\tablecomments{The observed polarizations and redshifts of GRBs.}
\end{deluxetable}

\section{Polarization induced by cold electrons via Compton scattering \label{sec:setup}}

In the standard fireball model, a jet expands outwards from the central engine, and finally coasts with a large bulk Lorentz factor $\Gamma$.
In the most forward layer of the jet, the internal energy can be ignored, and the electrons are considered to be non-relativistic in the  comoving frame.
In the inner part of the jet, the internal shock accelerates  electrons, leading to the SR in the presence of  magnetic fields. Therefore, the synchrotron
photons will illuminate the cold electron layer and be scattered by electrons via Compton processes. This process is different from the upper scattering Compton
process, since the latter transfers the energy from electrons to photons, while the former transfers energy inversely.
The optical depth is given as $\tau \equiv n \sigma_T r /\Gamma=L \sigma_T/ 4\pi r m_p c^3 \Gamma^3$, where $L$ is the isotropic luminosity, $\sigma_T$ is the
Thomson cross section \citep{Chang:2012a}. For the typical Gamma-ray burst parameters, $L \sim 10^{52}$ erg $\cdot$ s$^{-1}$ and $\Gamma \sim 10^3$, $\tau$ is  less than $1$ in
 the region of $r> 10^{10}$ cm. Thus, one reasonable assumption is that one photon is scattered at most once by the electrons in the forward layer of the jet.

\begin{figure}
\centering
  \plotone{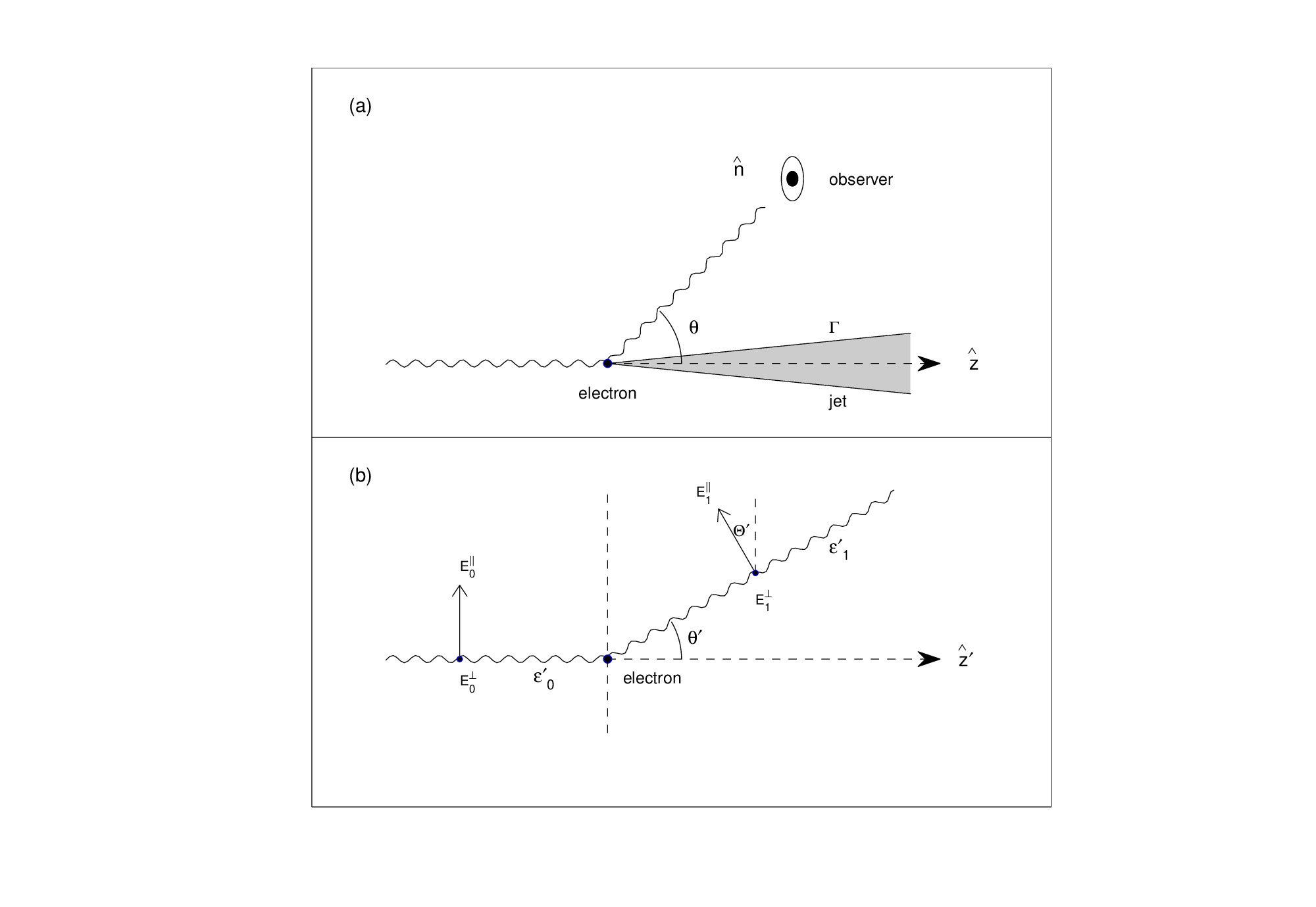}
  \caption{Schematic description of the CS processes. The top panel depicts an incident photon scattered in the observer frame.
The bottom panel describes the decomposition of electric vector before (E$_0$) and after (E$_1$) the scattering process in the comoving frame.
 The parallel direction ($\parallel$) and  transverse direction ($\bot$) are defined to be in  and orthogonal to the  scattering plane, respectively.} \label{fig:jet}
\end{figure}

 The set-up of this system is given in Figure \ref{fig:jet}.
The SR photons are collimated  along the moving direction of the outflow, namely the $\hat{z}$-direction, scattered off by a static electron
 at the ${\cal O}$ point, and then travel towards the observer in the $\hat{n}$-direction. The angle between $\hat{z}$ and $\hat{n}$ is
 $\theta$. The initial polarization is defined
 as $\Pi_0=(I'_{\parallel}-I'_{\bot})/(I'_{\parallel}+I'_{\bot})$, where the intensities $I'_{\parallel}$ and $I'_{\bot}$ correspond to the electric components
${\rm E}_0^{\parallel}$ and ${\rm E}_0^{\bot}$ (see Figure \ref{fig:jet}), respectively.
We use the convention that the variable with a prime is in the comoving frame.  Consider an incident photon with  energy $\varepsilon'_0$,
the photon energy after scattering in the comoving frame is given by
\beq \label{eq:scaene}
\varepsilon'_1= \frac{\varepsilon'_0}{1+\frac{\varepsilon'_0}{m_e c^2}(1-{\rm cos}\theta')}.
\eeq
The  cross section (the Kein-Nishina formula) reads
\beq \label{eq:crosec}
d \sigma  = \frac{r_0^2}{4} d \Omega' \left( \frac{\varepsilon'_1}{\varepsilon'_0}\right)^2 \left[ \frac{\varepsilon'_1}{\varepsilon'_0}
+\frac{\varepsilon'_0}{\varepsilon'_1}-2 +4 {\rm cos}^2 \Theta' \right], \eeq
where $r_0\equiv e^2/m_e c^2$ is the classical electron radius, and $\Theta'$ is the angle between the polarization direction before and after the scattering.
The polarization  is defined to be
\beq \label{eq:plodef}
\Pi' \equiv \frac{J'_{\parallel}-J'_{\bot}}{J'_{\parallel}+J'_{\bot}},
\eeq
where the intensities $J'_{\parallel}$ and $J'_{\bot}$ are written as
\begin{align} \label{eq:intpa}
J'_{\parallel} d \Omega' &= I'_{\parallel} d \sigma \left( \Theta'=\theta'\right)+I'_{\bot} d \sigma\left( \Theta'= \pi/2 \right), \\
J'_{\bot} d \Omega' &= I'_{\parallel} d \sigma\left( \Theta'= \pi/2\right)+I'_{\bot} d \sigma \left( \Theta'= 0 \right), \label{eq:intb}
\end{align}
respectively. Substituting Equations (\ref{eq:crosec}), (\ref{eq:intpa}) and (\ref{eq:intb}) into Equation (\ref{eq:plodef}),
one obtains the polarization in the comoving frame
\beq \label{eq:polssc}
\Pi'(\theta')= \frac{\Pi_0(1+ {\rm cos}^2\theta')-{\rm sin}^2 \theta'}{\varepsilon'_1/\varepsilon'_0+\varepsilon'_0/\varepsilon'_1
-(1+\Pi_0) {\rm sin}^2 \theta'  }.
\eeq
Since the incident photons are of the synchrotron origin, the initial polarization is $\Pi_0=(p+1)/(p+7/3)$, where $p$ is the power-law index
 of the electron distribution ($N(\gamma')\propto \gamma'^{-p}$) \citep{Rybicki:1979}. The polarization is invariant under Lorentz
transformation \citep{Cocke:1972}. One has $\Pi'(\theta')=\Pi(\theta)$, where $\theta$ is the angle in the observer frame. The Lorentz
transformation of the angle $\theta'$  to the observer frame is expressed as
\beq \label{eq:angtra}
{\rm cos} \theta' = \frac{{\rm cos} \theta - \beta}{ 1-\beta {\rm cos} \theta},
\eeq
where $\beta \equiv \sqrt{1-1/\Gamma^2}$ is the velocity of the jet in unit of the light speed. The inverse transformation gives
$\cos \theta= (\cos \theta'+ \beta)/(1 +\beta \cos \theta')$, therefore $\theta \ll 1$ for $\beta\approx 1$.
 The Doppler effect leads to $ \varepsilon'_0={\cal D}^{-1}_0 \varepsilon_0$ and $\varepsilon'_1={\cal D}^{-1} \varepsilon_1$,
where ${\cal D}_0=1/\Gamma (1-\beta )$  and ${\cal D}=1/\Gamma (1-\beta \cos \theta)$ are the Doppler factors.
One has such relation
\beq \label{eq:e1e0}
 \Sigma  \equiv\frac{\varepsilon'_1}{\varepsilon'_0}= 1- \frac{\Gamma \varepsilon_1}{m_e c^2}(1-\cos \theta)(1+ \beta).
  \eeq
Since $\Sigma$ is positive, we should require $\varepsilon_1<m_ec^2/[\Gamma(1-\cos \theta)(1+ \beta)]\approx m_e c^2 /\Gamma \theta^2 $.
When $\theta$ goes to zero, this limit is not complete. The spectrum of the incident photons can be described by the Band function in most GRBs,
we use $\varepsilon_{\rm max}$ to denote the upper limit of the high energy band. Since the incident photons lose energy to the static electrons
 in the scattering processes, another limit is $ \varepsilon_1 < \varepsilon_{\rm max}$. Therefore, $\varepsilon_1$ must satisfy
\beq \label{eq:constraint}
\varepsilon_1< {\rm min} \{ \varepsilon_{\rm max}, m_e c^2 /\Gamma \theta^2 \}.
\eeq
 Based on Equations (\ref{eq:angtra}) and (\ref{eq:e1e0}), the polarization in the observer frame is written as
 \beq \label{eq:ploobs}
 \Pi(\varepsilon_1, \theta) = \frac{\Pi_0[(1+{\rm cos}^2\theta)(1+\beta^2)-4\beta {\rm cos} \theta]- (1-\beta^2)\sin^2 \theta}{(1-\beta
{\rm cos} \theta)^2 (\Sigma+\Sigma^{-1})- (1-\beta^2)(1+\Pi_0){\rm sin}^2\theta},
 \eeq
 where $\Sigma$ is given in Equation (\ref{eq:e1e0}). Therefore, the observed polarization
 is a function of both the observation angle $\theta$ and the observed photon energy $\varepsilon_1$.

\begin{figure}
\centering
  \plotone{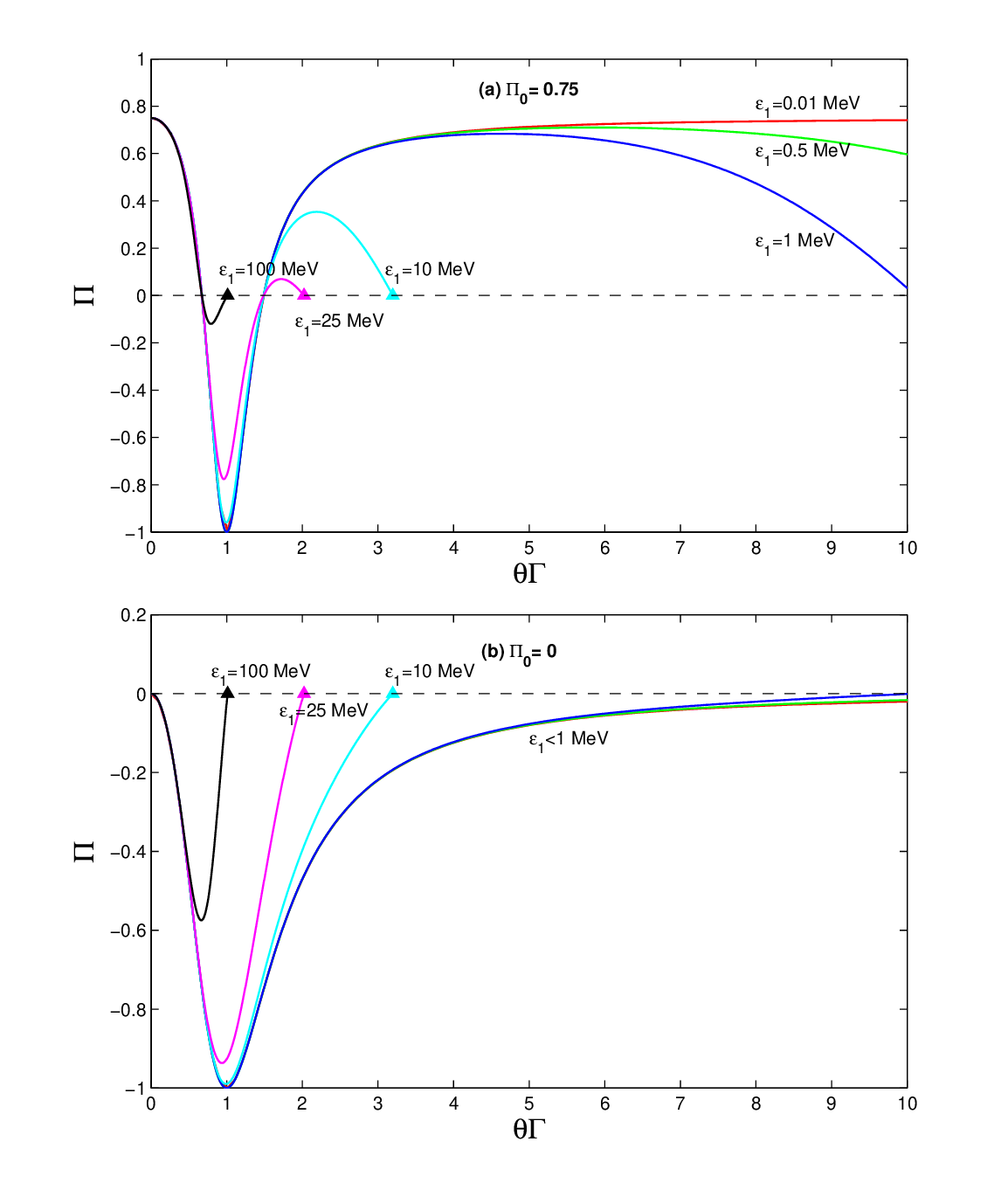}
  \caption{Polarization as a function of the observing angle. The initial polarization is set to be $\Pi_0 =0.75$ in the top panel, and
 $\Pi_0=0$ in the bottom panel. Polarization curves for photons with various energies $\varepsilon_1=0.01, 0.5, 1, 10, 25$ and $100$ MeV are
 plotted. }  \label{fig:p075}
\end{figure}

In Figure \ref{fig:p075}, we present the plots of the observed polarization at different viewing angles in the observer frame.
The shape of the curves depends on the  bulk Lorentz factor of the GRB outflow. For a typical long GRB, \citet{Chang:2012b} showed that
 the bulk Lorentz factor is approximately 200. Thus, we set $\Gamma=200$ in the numerical calculation. The initial polarization is set
to be $\Pi_0=0.75$ for the top panel, which is a typical value of the SR induced polarization.  For small viewing angle, i.e.
$\Gamma \theta \ll 1$, $\Pi$ is roughly the same with $\Pi_0$ for photons with different energies. When $\Gamma \theta \sim 1$, the polarization
direction changes $90^{\circ}$ and the polarization approaches $100\%$ for small energy photons, i.e., $\varepsilon_1 \leq 1$ MeV. As the energy
 increases, the polarization starts to decrease at this special viewing angle. For instance, $\Pi< 20 \%$ for $\varepsilon_1>100$ MeV.
For $\varepsilon_1=0.01$  MeV, $\Pi$ goes back to the initial $\Pi_0$  in large viewing angles,
and also the polarization direction returns.  However, for the
$\varepsilon_1=1$ MeV, $\Pi$ decreases to zero at $\Gamma \theta  \sim 10$. The regime in the parameter space of $\varepsilon_1=1 $ MeV and
 $\Gamma \theta >10$ is excluded by the constraint in Equation (\ref{eq:constraint}),  we cut off the curves for high energy photons.
When the energy of photons is 10 MeV, the polarization will be less than $40\%$
at the second peak. Even at $\Gamma \theta\sim 1$, the polarization starts to decrease for $\varepsilon_1=25$ MeV. The curve of 100 MeV
shows that the polarization almost shrinks to 0 for $\Gamma \theta =1$.

If the initial light beam is unpolarized, i.e. $\Pi_0=0$, the CS process can still cause polarization \citep{Shaviv:1995,Lazzati:2004,Toma:2009}.
 In the bottom panel of Figure \ref{fig:p075}, the polarization as a function of different viewing angles for different energies is plotted. For photons
 with energy less than 1 MeV, $\Pi$ gets the maximal value $100\%$ at $\Gamma \theta =1$, and approaches to zero quickly when $\Gamma \theta >5$. This is the
result of the Compton induced polarization in the Thomson limit. Thus,  one can observe completely polarized
gamma-rays at viewing angle $\theta \sim \Gamma^{-1}$ in the prompt phase. However, the polarization of $10$ MeV photons decrease to zero at $\Gamma\theta\sim 3$.
 The reason is that the Klein-Nishina effects start to influnce the polarization for high energy photons. For  $\varepsilon_1=100$ MeV, $\Pi $ decreases
to $60\%$ at the first ``valley", but goes to $0$ at $\Gamma\theta \sim 1$. The similar curves emerge in both panels, which can be attributed to
 the Klein-Nishina effects. In both $\Pi_0=0$ and $\Pi_0=0.75$ cases, the polarization of  high energy photons is smaller than that of low energy photons. This
 characteristic  is different from the polarization caused by the SR, and can be used to distinguish the polarization origin.

\begin{figure}
\centering
  \plotone{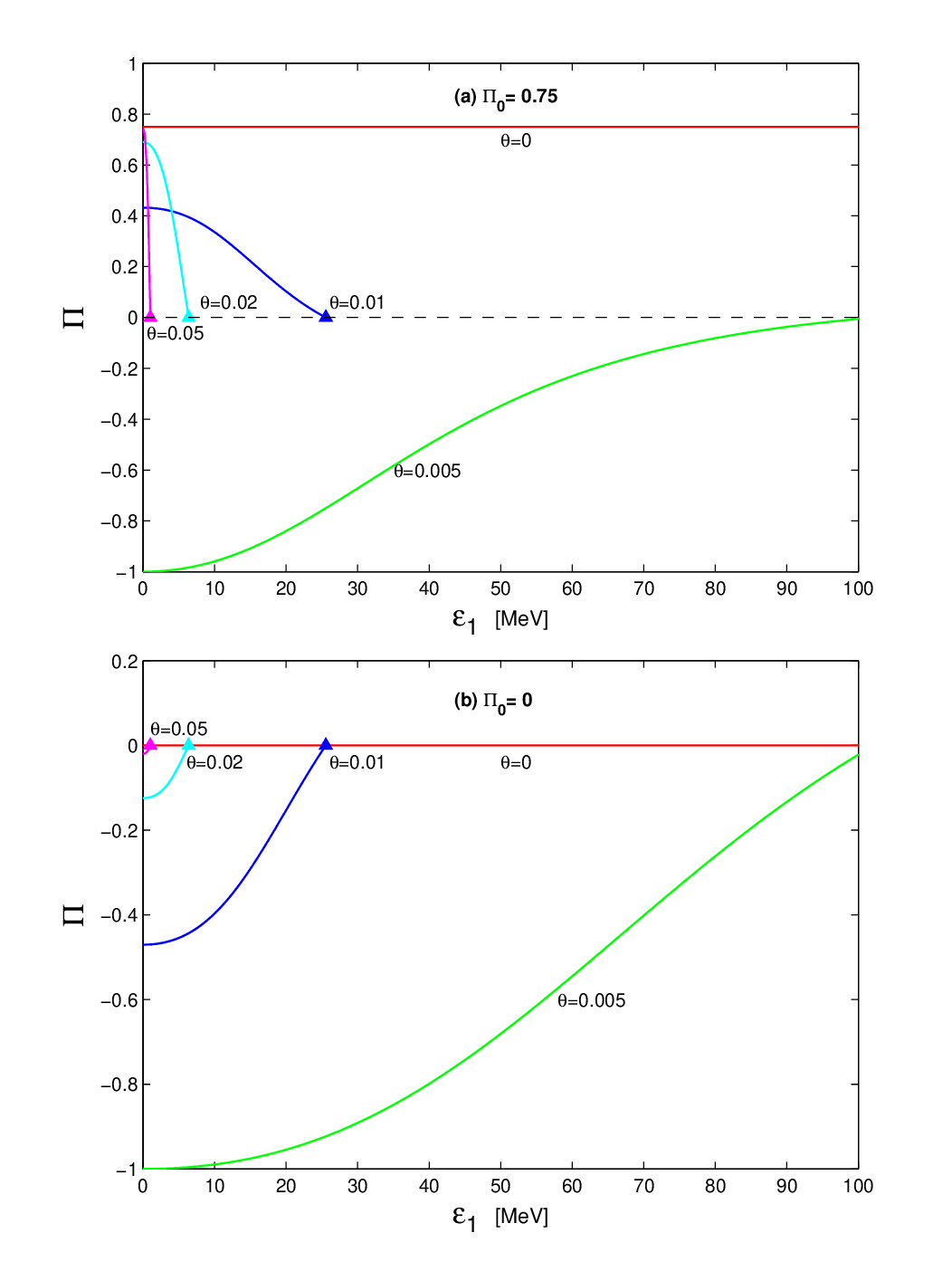}
  \caption{Polarization as a function of the observed photon energies. The initial polarization is set to be $\Pi_0 =0.75$ in the top panel, and $\Pi_0=0$ in
 the bottom panel. Polarization curves for different viewing angles $\theta=0, 0.005, 0.01, 0.02$ and $0.05$  are plotted.}  \label{fig:pe}
\end{figure}

From the observational point of view, the polarization as a function of photon energy for different fixed viewing angles is more convenient.
Figure \ref{fig:pe} depicts $\Pi(\varepsilon_1)$ for different viewing angles.  In the top panel, $\Pi_0=0.75$. For $\theta=0$, one can always observe
a uniform $75\%$ polarization for all photons. For $\theta=0.005$, which corresponds to the first ``vally" in the top panel of Figure {\ref{fig:p075}},
the polarization gradually goes to zero as the energy increases to $100$ MeV. For $\theta=0.01\sim 2/\Gamma$, $\Pi$ starts from $40\%$ for $\varepsilon_1 < 1$ MeV
and asymptotically goes to zero. For $\theta=0.02$ and $0.05$, the polarization degree goes quickly to zero for photons with energy less than $10$ MeV, and the polarization
 direction does not change.
In the bottom panel, we set $\Pi_0$ to be zero. No scattering happens for $\theta=0$, so no polarization will be observed.
 For $\theta=0.005$, which corresponds to the maximal polarization induced by CS.
$\Pi$ goes to zero when $\varepsilon_1$ varies from 1 MeV to about 100 MeV.

\begin{figure}
\centering
  \plotone{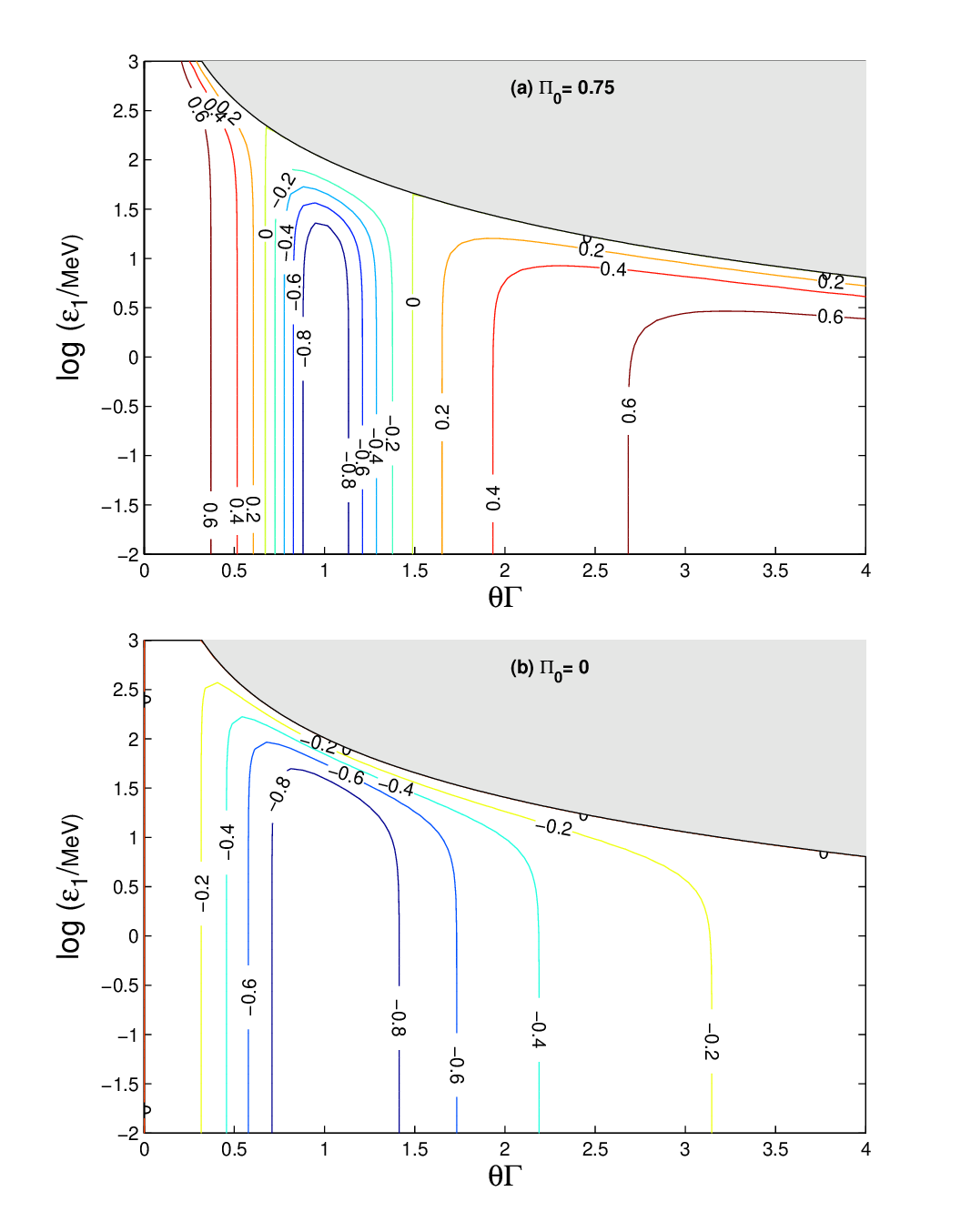}
  \caption{The contour representation of the polarization for $10$ keV $ < \varepsilon_1 < 1$ GeV and $0< \Gamma \theta <4$.
We set $\Pi_0=0.75$ in the top panel and $\Pi_0=0$ in the bottom panel. }  \label{fig:cm}
\end{figure}

A contour representation of the polarization is shown in Figure \ref{fig:cm}, the energy band is  $10$ keV $ - 1$ GeV and $\Gamma \theta$
is in the range $0 - 4$. The top panel depicts the SR induced initial polarization, and the bottom panel describes that the initial beam
is unpolarized. \citet{Toma:2009} calculated the polarization in the Compton drag model by using the stokes parameters, and showed that the maximal polarization
degree occurs when $\theta\sim 1/\Gamma$. The similar results are also obtained by \citet{Lazzati:2004}. In Figure \ref{fig:cm}, we show that high level
polarization occurs in both the initially polarized and unpolarized cases. In the top panel, the initial polarization can be observed for
$\Gamma \theta \ll 1 $, and there is no limit on the photon energy. The second large polarization occurs for $\Gamma \theta \sim 1$ and $\varepsilon_1 < 20 $ MeV,
 very hard gamma-rays are unpolarized at this view angle. However, the probability to detect this regime is strongly suppressed by the small cross section.
 The third high polarized regime is $\Gamma \theta >3$ and $\varepsilon_1 < 0.1 $ MeV, the maximal polarization is roughly $75\%$, this is predicted by the
 CS process in the Thomson limit. The polarization is negligible at the region of $\Gamma \theta> 2 $ and
$\varepsilon_1>20$ MeV. The bottom panel in Figure \ref{fig:cm} shows the polarization in the $\Pi_0=0$ case. One can also see the high polarized  regime,
where $\Gamma \theta \sim 1$ and $\varepsilon_1<20$ MeV. The grey area shows the forbidden parameter region.

\begin{figure}
\centering
  \plotone{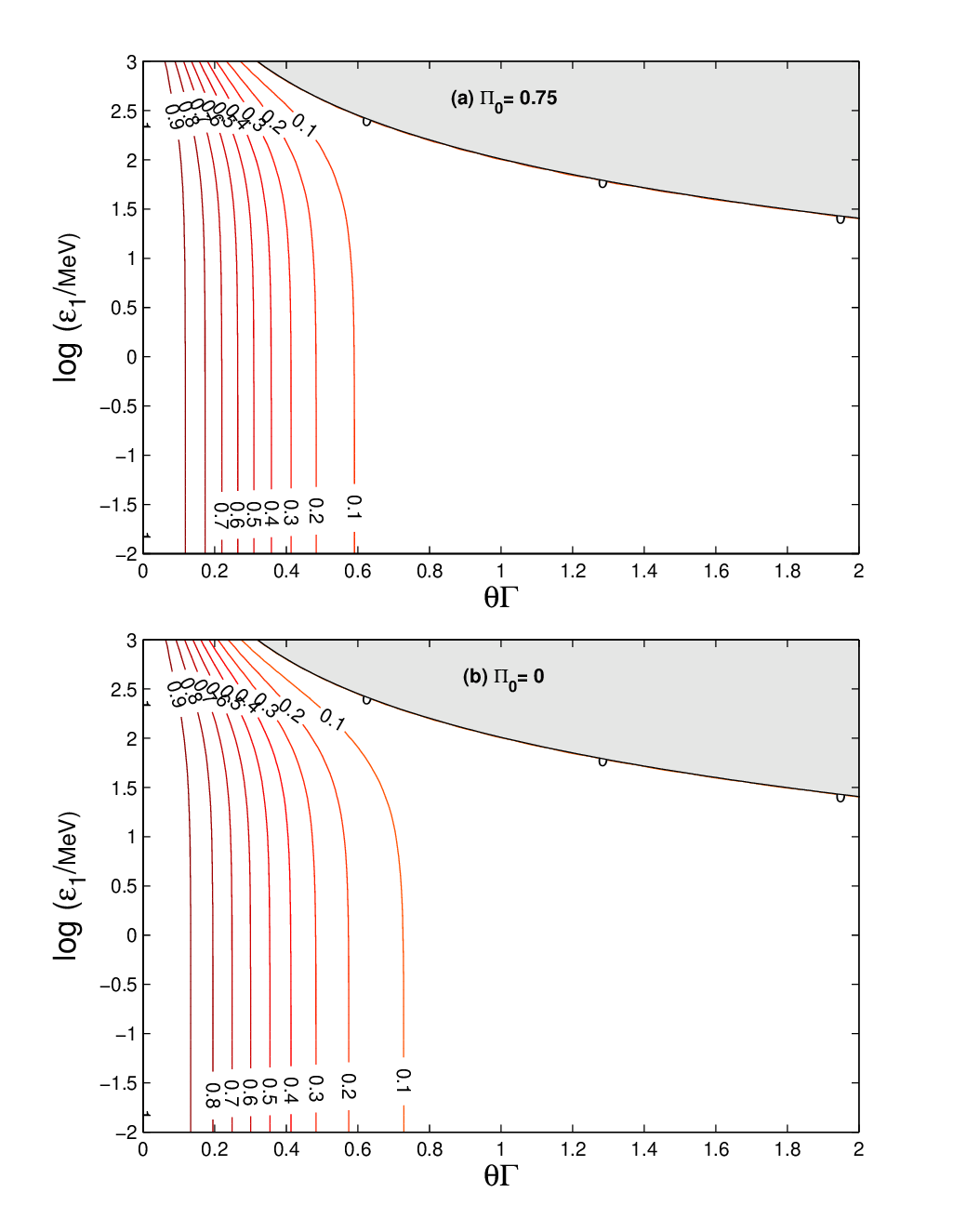}
  \caption{The contour representation of the intensity  in unit of  $r_0^2 I_0/ \sigma_{\rm total} $. We set $\Pi_0=0.75$ in the top
panel and $\Pi_0=0$ in the bottom panel.}\label{fig:cs}
\end{figure}

 The total intensity after scattering is $J'=J'_{\parallel}+ J'_{\bot}$, where $J'_{\parallel}$ and $J'_{\bot}$ are given in Equations (\ref{eq:intpa}) and (\ref{eq:intb})
\footnote{Here we ignore a normalization factor, i.e., the total cross section $\sigma_{\rm total}$.}. Under Lorentz transformation,
the received intensity transforms as $J = {\cal D}^4 J'$ \citep{Rybicki:1979}. The incident intensity $I'$
 transforms as $I'={\cal D}_0^{-4} I_0$, where  $I_0$ is  the observed intensity at $\theta=0$.
 Using Equation (\ref{eq:crosec}), we write the scattered intensity in the observer frame as
\beq \label{eq:labcs}
J = \frac{r_0^2}{2 \sigma_{\rm total}} I_0 \left( \frac{1-\beta}{1-\beta \cos \theta}\right)^4 \Sigma^2\left[ \Sigma+ \Sigma^{-1} - (1+\Pi_0)
\frac{\sin^2 \theta}{\Gamma^2(1-\beta \cos \theta)^2}\right].
\eeq
In Figure \ref{fig:cs}, we used the contour representation to express the intensity as a function of the energy and scattering angle. The intensity decays rapidly to zero
 when $\Gamma\theta$ is asymptotic to $1$, which means that the probability to detect the high polarization is strongly suppressed.

We  mainly discuss the polarization induced by the CS process between gamma-rays and cold electrons. Such Compton processes do
not change the original photon spectra significantly.  The first reason is that the optical depth is less than one in this regime. The typical
peak photon energy of the spectra is about $200$ keV in the observer frame. It is $1$ keV in the comoving frame if the bulk Lorentz factor
of the jet is $\Gamma\sim 200$.  Hence, photons in the prompt phase satisfy $\varepsilon'_0 \gamma'/m_e c^2 \ll 1$ ($\gamma'=1$ is the Lorentz
factor of cold electrons in the comoving frame), and the CS
is in the Thomson limit. Therefore, the photon energy is almost the same before and after scattering. For several hundred MeV photons, the head
 on collision will transfer the majority of the photon energy to the  electron, so the CS processes will reduce the number of high
 energy photon events. This may be the reason for that the LAT (One instrument on the Fermi satellite which detects the energy range from $30$ MeV to $300$ GeV)
 photons are only observed in few GRBs. It has been argued that the Compton induced polarization requires a large energy budget \citep{Coburn:2003}.
If the scattered photons are isotropic
in the comoving frame, the observed gamma-rays take only a fraction $d\Omega'/4 \pi$ of the total emission. However, this is not the case in the system
 of the present paper due to the Lorentz effects.  Although the angle distribution of scattered beam is nearly isotropic in the comoving frame,
 the scattered beam is emitted in a  small cone  in the observer frame.

One attractive feature of the Compton processes induced polarization is that the polarization can be as high as $100\%$. In the SR induced case,
the polarization depends on both the configuration of the magnetic fields and the electron spectral index $p$. Considering a uniform magnetic
 field at large scale with the most ideal configuration, $\Pi_{\rm max}=(p+1)/(p+7/3)<80\%$ for $p<4$. The shock accelerated electrons have a typical
 spectral index $p=2.3$, this predicts $\Pi_{\rm max}=71 \%$. Therefore, the CS process is the most probable to account for the
observation of $\Pi> 80 \%$. One can not exclude the SR as the major emission mechanism in the prompt phase by observing $\Pi> 80 \%$,
since $\Pi= 100\%$ at $\Gamma \theta=1$ occurs in both the $\Pi_0=0$ and the $\Pi_0=0.75$ cases. Also at this right viewing angle, the polarization
 direction changes $90^{\circ}$ relative to the incident polarization direction.

The analysis of the intensity also indicates that the scattered beam is relatively dimmer at $\Gamma \theta=1$. This means that the chance to detect the high
 polarization is small. However, this difficulty can be overcame with certain specific configurations. \citet{Levinson:2004} considered the polarization
 caused by gamma-rays scattered off the coasting baryon walls by the Compton sailing. There is a thick baryon wall in the out layer of the jet,
 and the inner core of the jet is baryon poor. This leads to the coincidence of  the angle of the maximal polarization  and  the angle of the maximal intensity.
 If we consider the similar wall sheath configuration, but replace the baryon materials
 with the cold electrons, the viewing angle coincidence of the maximal intensity and the maximal polarization can be realized.
Our derivation of the polarization is based on the point source, it still holds when the emission regime of the jet is a small surface. In the narrow jet
 scenario with $ \Gamma \theta_{\rm jet}< 1$, our results in Equation (\ref{eq:ploobs}) roughly holds.  When the emission surface is large, i.e.
$\Gamma \theta_{\rm jet} > 1$, the maximal polarization becomes small \citep{Lazzati:2004}.

 The polarization of SR photons is independent of the energy, but the CS induced polarization does depend on the energy.  If $\Gamma$ is larger, for instance $\Gamma=1000$,  $2$ MeV photons have smaller polarization than $1$ MeV photons.
The broad energy band detection of polarization is necessary to verify the CS induced polarization. This gives us a trend that higher energy photons
have smaller polarization, which distinguishes from that of the SR induced polarization.

An observer collects radiations from different emission regions with different viewing angles. The averaged polarization should consider
such net polarization effect. The geometry of the scattering region is assumed to be a plane, and the integral range depends on the size of the
plane. Other structures of the jet are also possible, which can be studied case by case. For simplicity, we set $\varepsilon_1=0.1$ MeV, which is the
typical value of the observed gamma-rays. Both the polarization and the intensity are only functions of $\Gamma \theta$.
The averaged polarization can be written as
\beq \label{eq:avepol}
\langle \Pi \rangle = \frac{\int^{x_2}_{x_1} \Pi(x) J(x) dx}{\int^{x_2}_{x_1}  J(x) dx},
\eeq
where $x \equiv \Gamma \theta$ and the integration range is from $x_1$ to $x_2$. Since the intensity  $J$ is almost zero when $x \sim 1$, we take $x_1=0$ and
$x_2=1$. For $\Pi_0=0.75$, we obtain $\langle \Pi \rangle\approx 65\%$. Hence, the CS process reduces the initial polarization.
For $\Pi_0=0$, we obtain $\langle \Pi \rangle\approx 20\%$. For randomly oriented magnetic fields with  N patches, the initial SR induced polarization
degree is  $\Pi_0 =\Pi_{\rm max}/ \sqrt{N}$, which can change $\langle \Pi \rangle$ significantly.
If the jet opening angle is small, i.e. $\theta_{\rm jet} \ll \Gamma^{-1}$,
the integral range becomes small. In this way, the averaged polarization is mainly determined by the specific viewing angle.
The probability of the CS process is relative to the optical depth $\tau=\sigma_T n R$, where the number density of cold electrons $n$ is an unknown parameter.
If $\tau <1$, one should also consider the mixing effect of the scattered and un-scattered  photons.
Considering all these ingredients, the SR plus CS model predicts a wide range of the polarization.

\section{Against  GRBs cases \label{sec:casestudy}}

We mainly concentrate on the polarization analysis in the prompt emission. \citet{Coburn:2003} reported a very large linear polarization
$\Pi=80\%\pm 20\%$ in the prompt phase of GRB 021206 observed by {\it RHESSI}. However, other independent groups did not confirm the polarization signals using the
 same data \citep{Rutledge:2004, Wigger:2004}. Similar debates also occured in GRB 041219A \citep{ Kalemci:2007,McGlynn:2007,Gotz:2009}. The instrumental systematics
 are the main obstacle to get a convincing result \citep{Yonetoku:2011}. In the following, we will discuss polarizations in
GRB 041219A \citep{McGlynn:2007}, GRB 100826A \citep{Yonetoku:2011}, GRB 110301A and GRB 110721A \citep{Yonetoku:2012}.

GRB 041219A is an intense burst detected by the {\it INTEGRAL}. The polarization measurement in the
 brightest 12 s interval and the total 66 s interval was performed in the energy ranges $100-300$ keV, $100-500$ keV and $100$ keV$-$1 MeV, respectively.
 With the 6 scattered directions analyses,  the polarization of the 12 s interval  is $98^{+2}_{-53}\%$ in the range $100-350$ keV and $71^{+29}_{-53} \%$ in the
 range $100-500$ keV \citep{McGlynn:2007}. Furthermore, the 3 directions analyses of the 12 s interval shows the polarization
to be $96^{+39}_{-40}\%$, $70 \pm 37 \%$ and $68\pm29\%$ in the three respective energy ranges. The Compton processes
predict  that the polarization of high energy photons is smaller than that of $E <1 $ MeV photons at $\Gamma \theta \sim 1$, and the critical energy is $10$ MeV.
 In order to lower the critical energy to 1 MeV, one needs  $\Gamma$ up to $2000$. The high luminosity of the burst means that the viewing angle is
not likely to be $\theta \sim \Gamma^{-1}$, unless the geometry structure of the jet is similar to the  baryon-wall structure given in \citet{Levinson:2004}. Beyond the
two conditions, the high polarization is most probably caused by the SR. Since the system uncertainty is large,
 a confirmative conclusion is difficult to give.

\citet{Yonetoku:2011} reported the polarization of GRB 100826A measured by {\it IKAROS}. The average polarization is $27 \% \pm 11\%$ in the energy range
$70 - 300$ keV with $2.9 \sigma$ confidence level. A change of polarization angle during the prompt phase was confirmed with $3.5 \sigma$ confidence level.
 GRB 100826A is the top $1\%$ of the
brightest events listed in the {\it BATSE} catalog. The peak energy is $E_{p}=606^{+134}_{-109}$ keV, and the low and high energy band indices are
$\alpha=-1.31^{+0.06}_{-0.05}$ and $\beta=-2.1^{+0.1}_{-0.2}$, respectively. This spectrum can be explained by the synchrotron radiation both
in the fast and the slow cooling phases. The data of the two intervals, each with the duration  of 50 s,  give  $\Pi_1=25\%\pm 15\%$ with $\phi_1=159 \pm 18$ deg
for Interval 1 and $\Pi_2=31\pm 21\%$ with $\phi_2=75 \pm 20$ deg for Interval 2, respectively. The pulse in  Interval 1 is  more intense than that in Internal 2.
With the SO model, \citet{Yonetoku:2012} argued that the polarization angle change in GRB 110826A is due to many patches of the magnetic
fields. The angle size of the magnetic fields satisfies $\theta_p \ll \theta_{\rm jet}$, where
$\theta_{\rm jet}$ is the jet opening angle. One can only observe an angle size of $\Gamma^{-1}$ along the line of sight. If $ \theta_{\rm jet}\sim \Gamma^{-1}$, many pathes can be observed,
and the polarization angle change is possible.
However, this can not explain why the polarization angle change is $90^{\circ}$ exactly.

If the initial jet openning angle is small, the polarization angle changing can be explained by the changing of the viewing angle in the
SR plus CS model.
 During the first interval, the line of sight moves away from near the axis of the
jet. In the range $1/2< \Gamma \theta < 2/3$ ($\Pi=0$ when $\Gamma \theta \approx 2/3$),  the SR induced polarization is dominated, and the average polarization
 is estimated to be $28\%$ according to Equation (\ref{eq:avepol}). This value is positive, referring the top panel in Figure \ref{fig:p075}. In the range
$2/3 < \Gamma \theta < 1$, the  averaged polarization  is $40\%$ with a minus sign, i.e., the polarization angle changes $90^{\circ}$.
Also the intensity in the second viewing angle range is smaller than the intensity in the first one. Then beyond $\Gamma \theta >1$, the intensity almost disappears.
And correspondingly, no respective pules is reported in observation. One can also infer that  $\theta_{\rm jet} \sim \Gamma^{-1}$ in this burst.
Therefore, the averaged polarization, the polarization angle and the light curves of the GRB 100826A can be well described in the SR plus CS model .

Finally, we discuss the recent polarization measurements in GRB 110301A and GRB 110721A \citep{Yonetoku:2012}. Compared with GRB 100826A, no polarization angle change
was detected in these two bursts, the polarization is $\Pi=70\% \pm 22\%$ ($3.7$ $\sigma$) for GRB 110301A and $\Pi=84^{+16}_{-28} \%$ ($3.3$ $\sigma$) for
GRB 110721A. \citet{Yonetoku:2012} explained that the synchrotron model can be  consistent with these two GRBs, the magnetic field structures
are globally ordered and advected from the central engine. GRB 110301A has a short time duration, $T_{90}=5$ s, and the peak energy is about  $E_{\rm peak} = 106.80 $
keV \citep{Foley:2011}.
The Band spectra gives $\alpha = -0.81$ and $\beta = -2.70$. One can obtain the index of electrons via the relation $\beta=-(p+1)/2$, i.e. $p=4.4$. The SO model
predicts $\Pi=80\%$, consistent with the observation. GRB 110721A has an unusual high energy peak $E_{\rm peak} \sim 15 $ MeV \citep{Axelsson:2012}, the dissipative
photosphere synchrotron model can account for such high peak values whether the outflow is extreme magnetic-dominated or baryon-dominated \citep{Veres:2012}.
The gamma-ray burst polarimeter  aboard the {\it IKAROS} mainly observes the energy range $70-300$ keV, which coincides with the blackbody component with
temperature ranging in  $10-100$ keV. However, the flux of the blackbody component is small compared to the non-thermal component, and the polarization reduction is less
important. The low energy index of the Band spectra is $\alpha\approx -1$, which is the mostly observed value in GRBs.  The SR in the fast
cooling phase predicts $\alpha \sim -2/3$. Some authors suggested that the IC  scattering in the Klein-Nishina  regime plays an important role  to tune
 $\alpha$ to approach the $-1$ limit \citep{Nakar:2009,Duran:2012bk}. The IC processes in the hot plasma will reduce the polarization to a neglectful level. Meanwhile,
the energies of the scattered photons are also shifted to the high energy range $E >1$ MeV. The observed photons in the  energy range $70-300$ keV are due to the low
energy tail of the SR. The observed high polarization can not exclude the self-synchrotron Compton (SSC)  origin.

\section{Conclusions and remarks \label{sec:con}}

We showed that the synchrotron photons collide with cold electrons in the jet can significantly change both the polarization degree and the
polarization direction. The photons are not up-scattered, but transfer energies to electrons. Due to the Klein-Nishina effects, high energy photons ($E>10$ MeV) have
smaller polarization than low energy photons ($E< 1$ MeV). After scattering,
 the polarization angle changes $90^{\circ}$ relative to the original direction, and the polarization reaches the maximal value at the right viewing angle $\Gamma \theta \sim 1$.
These results indicate that the high polarization may also caused by the CS process. The jet structure is essential for the net polarization, we leave
this topic for future study.

The polarimetry of the prompt emission can be used to distinguish the GRB models. In the SO model,
the prompt emission is due to the synchrotron radiation, an ordered magnetic fields can produce large polarization. The SO model predicts that
the polarization is universal for photons with different energies, the unknown geometry of the emitting region also affects the observed results \citep{Granot:2003, Nakar:2003}.
Recently, the deviations from the Band function in the low energy range have been discussed by \citet{Tierney:2013}.
For instance, a Band plus blackbody fit is better than a single Band fit in the GRB 090323. The thermal component occurs naturally in
 the photosphere internal shock model \citep{Toma:2011}. The thermal photons have no polarization initially. After up-scattered by the electrons
 with power-law distribution via IC processes, the observed photons are linearly polarized. The linear polarization degree is anticorrelated with
the weight of the thermal component. Therefore, the polarization in the low energy range is less than that in the high energy range. This prediction
is quite different with the prediction in our SR plus CS scenario. The high polarization
can be obtained from the edge of a narrow jet, although the probability is not high \citep{Fan:2009}\footnote{
Our SR plus CS scenario is similar to Fan's set-up except two differences. One difference is that the incident photons are polarized
 in our scenario and not polarized in Fan's scenario. The other difference is that we considered cold electrons, while Fan considered
 the electrons with isotropic distribution. We leave the study the polarization induced by the polarized incident photons and isotropic electrons to be
the near future work.}.
\citet{Zhang:2011} proposed the internal-collision-induced magnetic reconnection and turbulence (ICMART) model
to explain GRB 080916C. The internal shock distorts the ordered magnetic fields, and further the magnetic turbulence triggers
 the prompt emission. One ICMART event corresponds to one pulse in the GRB light curve. The linear polarization degree evolves during one pulse.
In the beginning of one pulse, the polarization degree achieves the maximal value of SR, i.e., $\Pi \sim 50 \%-70 \%$. At the end of the pulse, the
ordered magnetic field structure is destroyed, and the reasonable net polarization is less than $10 \%$. The average value of the polarization in
 one pulse is estimated to be around $30 \%$, close to that observed in GRB 100826A. However, a physical scenario to explain the $90^\circ$ angle
changing in the two intervals is still missing. In summary, the polarization can help us a lot to understand the prompt emission.

%%%%%%%%%%%%%%%%%%%%%%%%%%%%%%%%%%%%%%%%%%%%%%%%%%%%%%%%%%%%%%%%%%%%%%%%%%%%%%%%%%%%%%%%%%%%%%%
\begin{acknowledgments}
We thank the anonymous referee for very helpful suggestions and comments. We are grateful to Li  M. H.,  Li X. and  Wang S. for useful discussion. This work has been supported  in part by the NSF of China under Grant  No.
11075166 and No. 11147176. Jiang Y. is also funded by  the China Postdoctoral Science Foundation funded project Grant No. 2012M510548.
\end{acknowledgments}

\end{document}